\documentclass[letter]{myaa} 

\usepackage{graphicx}
\usepackage{txfonts}
\usepackage{natbib}
\bibpunct{(}{)}{;}{a}{}{,} 
\newcommand{\mygi}{MyGIsFOS}

\def\teff{$T\rm_{eff}$}
\def\kms{$\mathrm{km\, s^{-1}}$}
\newcommand{\glog}{\ensuremath{\log {\rm g}}}

\usepackage{color}
\usepackage[normalem]{ulem}
\idline{10}

\begin{document}

   \title{Chemical abundances of giant stars in the Crater stellar system\thanks{based
on observations taken at ESO Paranal with the Kueyen telescope, programme 094.D-0547}
}

   \subtitle{}

\author{
P. Bonifacio \inst{1}
\and
E. Caffau \inst{1}\thanks{MERAC fellow}
\and
S. Zaggia  \inst{2}
\and
P. Fran\c{c}ois \inst{1,3}
\and
L. Sbordone \inst{4,5}
\and
S.M. Andrievsky\inst{6,1}
\and
S.A. Korotin\inst{6}
          }

\institute{ 
GEPI, Observatoire de Paris, PSL Research University, CNRS, Univ Paris Diderot, 
Sorbonne Paris Cit\'e  Place Jules Janssen, 92195 Meudon, France
\and
Istituto Nazionale di Astrofisica,
Osservatorio Astronomico di Padova, Vicolo dell'Osservatorio 5, 35122 Padova, Italy
\and
UPJV, Universit\'e de Picardie Jules Verne, 33 Rue St Leu, F-80080 Amiens
\and
Millennium Institute of Astrophysics, Chile
\and
Pontificia Universidad Cat{\'o}lica de Chile
Vicu{\~n}a MacKenna 4860, Macul, Santiago, Chile
\and
Department of Astronomy and Astronomical Observatory, Odessa National
University, and Isaac Newton Institute of Chile Odessa branch,  
Shevchenko Park, 65014 Odessa, Ukraine
}

   \date{}
 
  \abstract
{}
   {We obtained spectra for two giants of Crater (Crater\,J113613-105227 
    and  Crater\,J113615-105244) using X-Shooter at the VLT, with
    the purpose of determining their radial velocities and metallicities.}
   {Radial velocities were determined by cross-correlating the spectra with that
    of a standard star. 
    The spectra were analysed with the \mygi\ code using a grid of synthetic
    spectra computed from one-dimensional, local thermodynamic equilibrium (LTE) model 
    atmospheres. Effective temperature and surface gravity were derived from photometry
    measured from images obtained by the Dark Energy Survey.}
   { The radial velocities are $144.3 \pm 4.0$ \kms\  for Crater\,J113613-105227 and
    and $134.1 \pm 4.0$ \kms\ for Crater\,J113615-105244.
    The metallicities are [Fe/H]=--1.73 and [Fe/H]=--1.67, respectively.
    In addition to the iron abundance, we were able to determine abundances for nine elements: 
    Na, Mg, Ca, Ti, V, Cr, Mn, 
    Ni, and Ba. For Na and Ba we  took into account
    deviations from LTE because the corrections are significant. 
    The abundance ratios are similar in the two stars and resemble
    those of Galactic stars of the same metallicity.
    In the deep photometric images we detected several stars that lie
    to the blue of the turn-off.
}
   {The radial velocities imply that both stars are members of the
     Crater stellar system.  The difference in velocity between the
     two  taken at face value 
     implies a velocity dispersion $> 3.7$ \kms at a 95\% confidence level.
     Our spectroscopic metallicities agree excellently well with
     those determined by previous investigations using photometry. Our
     deep photometry and the spectroscopic metallicity imply an age of
     7\, Gyr for the main population of the system. The stars to the
     blue of the turn-off can be interpreted as a younger population
that is      of the same metallicity and an age of 2.2\,Gyr.
     Finally, spatial and kinematical parameters support the idea that
     this system is associated with the galaxies Leo~IV and Leo~V.
     All the observations favour the interpretation of Crater as a dwarf galaxy.}
{}


   \maketitle
%
\begin{figure}
\begin{center}
\resizebox{\hsize}{!}{\includegraphics[clip=true]{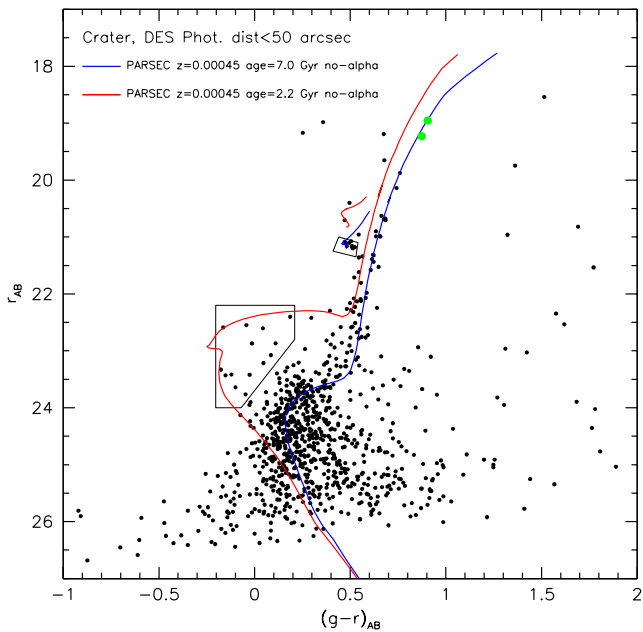}}
\end{center}
\caption[]{Colour-magnitude diagram in g and r bands of the Crater
  stars within 50\farcs{0} from the centre, based on the DES public
  images. The two target stars are identified as large green dots.
  Overplotted are shown two PARSEC isochrones for a 7 Gyr and a 2.2
  Gyr old stellar population.  A reddening correction
of E(B-V)=0.023, as deduced from the maps of \citet{Schlegel}, 
has been applied to the  isochrones.
\label{cmd}}
\end{figure}
\begin{figure}
\begin{center}
\resizebox{\hsize}{!}{\includegraphics[clip=true]{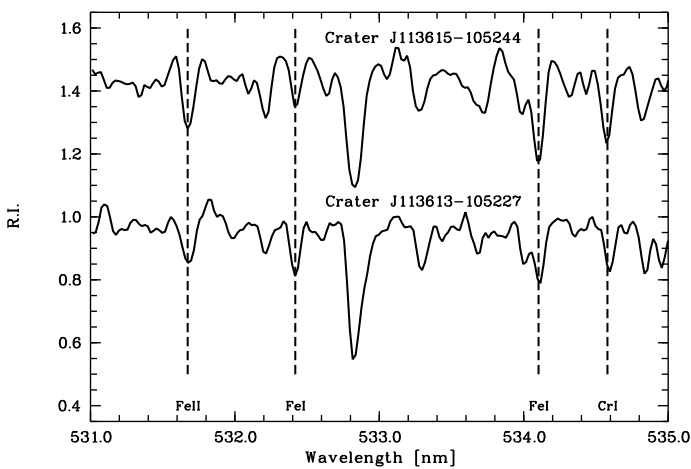}}
\end{center}
\caption[]{Portion of the X-Shooter  normalised spectra of the two stars,
R.I. is the residual intensity.
The spectrum of Crater\,J113615-105244 has been displaced
vertically  by 0.5  for display purposes. 
Only lines that have been used in the abundance analysis are identified.
\label{spectra}}
\end{figure}
\begin{figure}
\begin{center}
\resizebox{\hsize}{!}{\includegraphics[clip=true]{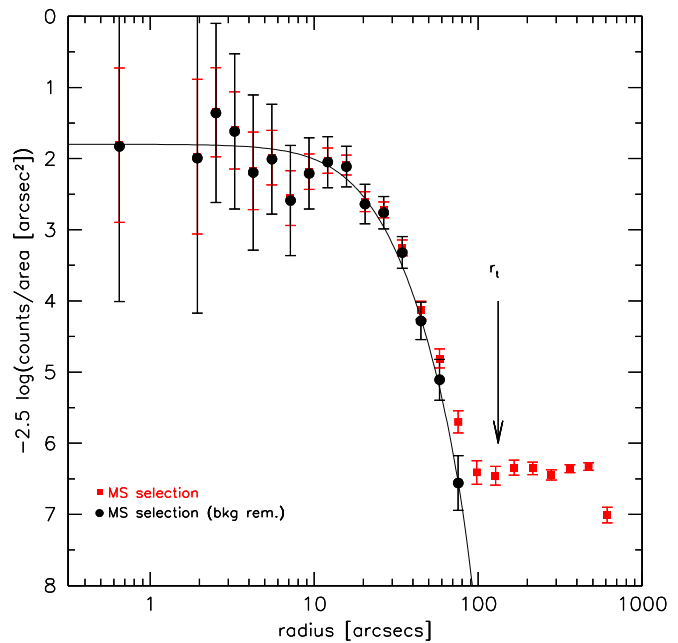}}
\end{center}
\caption[]{Radial profile of Crater. The red squares are the raw counts
of Crater stars selected along the stellar sequence; the filled circles
represent the counts corrected for foreground contamination using the
colour-magnitude diagram. The solid line is the best-fitting King
model \citep{King}, and the position of the tidal
radius is indicated by a downward arrow.
\label{prof}}
\end{figure}

\section{Introduction}

The compact stellar system Crater has been independently 
discovered by \citet{Belokurov2014}
from the ATLAS ESO VST survey \footnote{http://astro.dur.ac.uk/Cosmology/vstatlas/}
\citep{Shanks} supplemented with deep imaging with the WHT 4m telescope,
and by  \citet{Laevens2014} from the Pan-STARRS1  Survey
\footnote{http://pan-starrs.ifa.hawaii.edu/public/} \citep{Chambers},
supplemented by deep photometry with ESO/MPG 2.2m telescope.
\citet{Laevens2014} referred to the system as PSO J174.0675-10.8774, in
the present paper we adopt the naming proposed by
\citet{Belokurov2014}, which is more in line with the current convention of
naming new dwarf galaxies from the constellations in which they
reside: Crater.  The two investigations substantially agree
with respect to the distance of the system: between 145 kpc and 170
kpc for \citet{Belokurov2014} and 145 kpc $\pm 17$ kpc according to
\citet{Laevens2014}.  They also agree on the estimate of the
metallicity of the system: $<-1.8$ according to \citet{Belokurov2014},
$-1.9$ according to \citet{Laevens2014}.  However, the two
investigations depart significantly as to the nature of the stellar
system.  \citet{Belokurov2014} interpreted it as a dwarf galaxy
satellite of the Milky Way, although the hypothesis
that it is a peculiar globular cluster (GC) is also discussed.  
This interpretation is supported by the
existence of a few bright blue stars present within in the stellar
system, which can be interpreted as young blue-loop giants,
resulting from a recent star formation episode.  In addition,
in support of the dwarf galaxy interpretation, \citet{Belokurov2014} pointed out
that the horizontal branch (HB) of the system is quite red and not
very extended: such a morphology of the HB is not compatible with
low metallicity.  Low-metallicity GCs show HBs that are
very extended to the blue.  \citet{Laevens2014} instead interpreted the
system as a distant GC, for which they derived a precise
distance, metallicity, and age (8 Gyr). If this interpretation is
correct, then the bright blue stars observed in the system are either
blue stragglers or non-members.  It should be stressed that no other
such ``young'' metal-poor GC has been observed
to date. Such
a system could have been formed in a currently disrupted satellite
galaxy and now been accreted to the Milky Way halo.

In this Letter we present the result of spectroscopic observations
of two of the brightest giants of Crater. Our main motivation was
to determine radial velocities and metallicities for the two stars.

\section{Target selection, observations, and analysis}

In the two photometries presented by \citet{Belokurov2014} and
\citet{Laevens2014}, two luminous red giant stars are visible
at g$\approx 20$ at
$\simeq2$~mag over the HB. From our previous experience
with X-Shooter and the Exposure Time calculator, we estimated that the
stars were bright enough to obtain a
spectrum suitable for measuring radial velocities with X-Shooter
and determine
abundances in about four hours of exposure for each star.  We
retrieved {\it g, r} public images from the ESO\ archive and catalogues
of Data Release 1 of VST ATLAS and used them to identify the stars
and provide the coordinates used for the observation. Since magnitudes
in the catalogue were obtained with aperture photometry, we ran
DAOPHOT/ALLSTAR on the released images to extract PSF photometry to improve the coordinates and photometry of the targets, which
are already
at the detection limit.  The final photometry was corrected for
illumination using the ATLAS DR1 catalogue.

We subsequently retrieved public {\it g, r } deep images observed in
the course of the Dark Energy Survey (DES) \citep{Abbott,Diehl} from
the NOAO Science Archive\footnote{http://portal-nvo.noao.edu/}.  We
used two images, obtained from the stacking of several short
exposures, for each band for a total time of 6240 sec (3720+2520). The
seeing of the images is quite good $\simeq0.8\div0.9$ arcsec. We
extracted PSF photometry with DAOPHOT/ALLSTAR and used the multiple
exposures to obtain a deeper photometry with ALLFRAME. 
The final {\it g,r }catalogue was calibrated with the ATLAS DR1 photometry,
which is tied to the APASS system \citep{Henden,APASS}.
Magnitudes are AB magnitudes and should be
on the DECAM system.  
The DES adopted
photometry was used to create the {\it g,r} colour-magnitude diagram
of Crater as shown in Fig.\,\ref{cmd}, where the two observed stars
are highlighted in green.  Coordinates and photometry for our stars are
given in Table \ref{allstar}.  The two stars are quite centrally
located in Crater.
Our observing programme was to observe each star with
X-Shooter \citep{vernet} with the Integral Field Unit, which re-images
an input field of 4"x 1.8" onto a pseudo slit of 12"x 0\farcs{6}
\citep{IFU}.  Four observing blocks (OBs) of one hour each should have
provided 12136 s of integration on each star as well as a resolving power of
R=7900 \relax in the UVB arm and 12600 \relax in the VIS arm.
Four OBS on Crater\,J113613-105227
were executed, but only two OBs on Crater\,J113615-105244.
Although with a lower S/N than we had expected, the latter spectra turned
out to be of sufficient quality for our analysis.  The spectra were
reduced as described in \citet{gto11}.
The S/N ratio of the coadded spectra at 530\,nm was about 60 for
Crater\,J113613-105227
and about 40 for Crater\,J113615-105244. 
A portion of the spectra is shown in Fig.\ref{spectra}.

\begin{table}
\caption{\label{allstar}
Observing and measured parameters of our programme stars} 
\renewcommand{\tabcolsep}{3pt}
\tabskip=0pt
\begin{center}
\begin{tabular}{lrr}
\hline\noalign{\smallskip}
Parameter & J113613-105227 & J113615-105244\\
\noalign{\smallskip}\hline\noalign{\smallskip}
RA  (J2000.0) &  11:36:13.90 &  11:36:15.94 \\
Dec (J2000.0) & -10:52:27.40 & -10:52:44.10 \\
$r$    [mag]  & 18.954    &     19.230          \\ 
$(g-r)$  [mag]& 0.905     &      0.874          \\ 
E(B--V) [mag] & 0.023     &      0.023          \\ 
\noalign{\smallskip}\hline\noalign{\smallskip}
v$_{rad}$  [\kms]  & $144.3\pm 4.$ & $134.1\pm 4.$ \\
T${\rm eff}$ [K]          &  4575  &   4643  \\ 
\glog [cgs]               &   1.2  &    1.4  \\
$\xi$    [kms]            &   2.0  &    2.0  \\
${\rm [Fe/H]}$  [dex]     & --1.73 &  --1.67 \\  
${\rm [\alpha/Fe]}$ [dex] &   0.34 &    0.41 \\    
\noalign{\smallskip}\hline
\end{tabular}
\end{center}
\end{table}
\begin{table}
\caption{\label{size}
Results of the fit of a King model to the MS stars of Crater.} 
\renewcommand{\tabcolsep}{3pt}
\tabskip=0pt
\begin{center}
\begin{tabular}{lrr}
\hline\noalign{\smallskip}
Quantity & best fit & error \\
\hline\noalign{\smallskip}
core radius  & 26\farcs{5} & 4\farcs{3}\\
tidal rdius  & 132\farcs{5} & 15\farcs{2} \\
concentration & 0.70  & 0.12 \\
\hline\noalign{\smallskip}
\end{tabular}
\end{center}
\end{table}

\subsection{Photometry}

We used the DES deep photometry to measure the size of Crater. From
the colour-magnitude diagram we selected Crater stars brighter than
$r=25.0$ (i.e. $\simeq 1$ mag below the turn-off, TO) along the old
stellar sequence taking into account the local errors in colour, and
we used this sample to determine the centre of Crater. We excluded the
blue stars above the TO and counted the number of stars per unit area
in radial annuli as a function of distance from the centre. The
results of these counts are shown in Fig.\ref{prof}.
We fitted a King model
\citep{King} to these number counts, and the results of our best fit
are provided in Table \ref{size}. These compare well to what was determined
by \citet{Laevens2014} assuming a King model, although we 
point out that we derived a smaller core-radius and a larger
(almost a factor of 2) tidal radius. The reason is probably that DES photometry is deeper than the WFI photometry
of \citet{Laevens2014}.
The colour-magnitude diagram
shown in Fig.~\ref{cmd} for the stars $\le
50$\farcs{0} or $\le 1.9$ core radii  from the centre
confirms that the dominant population of Crater is  of intermediate age. 
The PARSEC
isochrones of the Padova Group \citep{parsec}  in the DECAM AB
system  for a metallicity of
[Fe/H]$=-1.5$ ($Z=4.5\times 10^{-4}$),  and  
reddened for the extinction
expected for Crater from the \citet{Schlegel} maps, 
are in favour of a younger age of $\simeq7$~Gyr compared
to the analysis of \citet{Belokurov2014}.  
We do not see a very young population of $\simeq 400$~Myr, but the good seeing and
cleaner photometry allow us to identify a population of objects
$\approx 1.5$ magnitudes above the turn-off region.  We visually
inspected these objects to verify that they are not background
compact galaxies or image artefacts, and they all appear to be stellar
objects. The younger population is compatible with an age of $\sim
2.2$ Gyr and the metallicity is compatible with the one of the main
population.  A population of
blue stragglers with a mass of $\simeq1.4$ M$_\odot$.
might account for some of the younger potential MS 
population, but the almost horizontal track of points heading toward the 
sub-giant branch does not follow an expected blue-straggler locus.
With our analysis we measured a distance modulus of
(M-m)$_0=20.91\pm0.1$ mag or a distance of 152 kpc, confirming the
distance found by the previous authors.

\subsection{Radial velocity \label{radvel}}

We measured the radial velocities given in Table \ref{allstar}
by cross-correlating 
each individual exposure in the UVB arm
against the X-Shooter spectrum of HD\,165195, taken
from the X-Shooter spectral library \citep{xsl}.  This star has
atmospheric parameters quite close to our programme stars.
We then applied  the barycentric correction to
the radial velocities and added a zero-point correction.
The zero-point correction was derived for each exposure
from the [OI] 577.7\,nm atmospheric emission line.
The zero-point corrections are probably due
to a combination of flexures in the spectrograph,
variation in temperature, and pressure in the surroundings
with respect to when the calibration Th-Ar was taken,
as well as centring of the star on the IFU.
The use of an emission line is not optimal for this
purpose, since the illumination of the slit is different
for the sky emission and for the star, but we do not
have strong enough telluric absorprion lines  to use
in the UVB arm. 
Interestingly, \citet{Schon} have performed
several measurements of the  [OI] 577.7\,nm line
at different positions on the slit on X-Shooter
slit spectra, and the dispersion of their measurements 
is of only 0.6 \kms.
We cannot make use of the VIS
spectra for the radial velocities for two reasons: since
the stars are metal-poor, there are few lines in the VIS range,
and the observations were made by tracking at 470\,nm, thus
the UVB spectrum is expected to remain well centred, while the
VIS spectrum probably shows larger deviations during
the exposure\footnote{See the X-Shooter manual-
page 27
https://www.eso.org/sci/facilities/paranal/instruments/xshooter/doc/VLT-MAN-ESO-14650-4942\_P96.pdf}.

Then we computed a weighted average of
the radial velocities, with weights that are proportional
to the number of counts in the spectrum in 
the region 490-550\,nm. The use of a straight average changes
the mean by less than 0.1 \kms.
 
\begin{figure}
\begin{center}
\resizebox{\hsize}{!}{\includegraphics[clip=true]{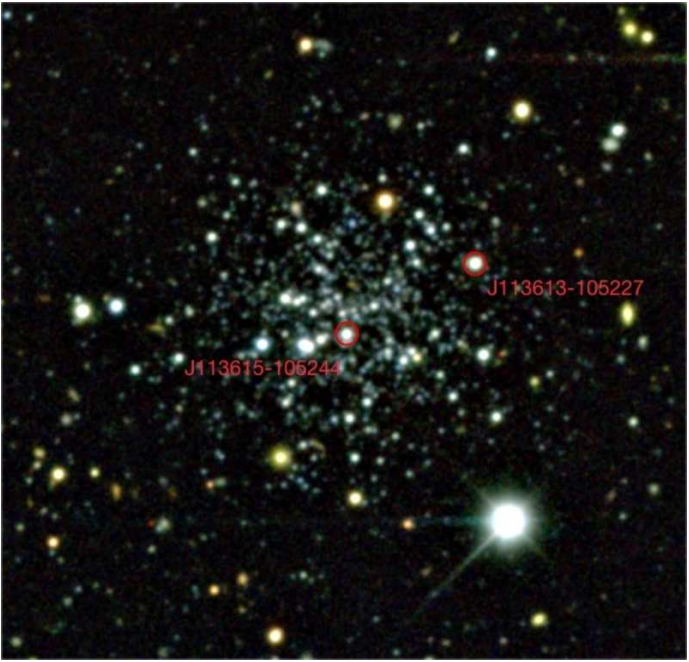}}
\end{center}
\caption[]{Colour image created from DES, g,r,i images showing the
  location of the two target stars.
\label{crater}}
\end{figure}

Measuring the radial velocities on the indivudal
spectra has the additional advantage that the dispersion
in the measurements provides an estimate of the error.
For Crater\,J113613-105227, this is 4 \kms , while  it is 0.4 \kms\ for
Crater\,J113615-105244.
We take the former as our error estimate, since
the spectra were taken  on three different
nights and probe different instrument flexures,
ambient temperature, and star centring, while the two spectra
of Crater\,J113615-105244  were taken on the same
night, one after the other. The latter do provide an estimate
of the statistical error in our radial velocity measurements.
Our error estimate  is in line with the quoted accuracy of the
calibration of the X-Shooter UVB and VIS arms
\citep[][2\,\kms]{Goldoni}. 
The difference in radial
velocity, $10.2$\kms, allows us to confirm that the two stars
belong to the same stellar system. This 
was expected because the stars are near the centre of Crater, 
as can be appreciated in Fig.\ref{crater}.
The average radial velocity of Crater is $139.2\pm 4.$ \kms.  
The radial velocity in the Galactic standard
of rest  is  $v(GSR)=-5.7\pm 4.$ \kms \ using the \citet{DB98} correction for the solar motion
and a correction for the rotation of the
Galaxy of $220.0$ \kms .

\subsection{Abundance analysis}
To analyse the spectra, we used our code \mygi\ \citep{mygisfos} and a
grid of synthetic spectra appropriate for metal-poor giants, computed
from our grid of ATLAS 12 models (Sbordone et al. in preparation,
\citealt{K2005,C2005}) covering the metallicity range $-3.0$ to
$-1.0$.  The spectra were computed assuming local thermodynamic
equilibrium (LTE).  
For both stars we adopted the atmospheric parameters
as read from the 7\,Gyr isochrone that is shown in Fig. \ref{cmd}.
The adopted surface gravites are also supported by the
iron ionisation equilibria, even though we have only two \ion{Fe}{ii}
lines in Crater\,J113613-105227 and one in  
Crater\,J113615-105244. 
\mygi\ is also  capable of determining the temperature
from the \ion{Fe}{i} excitation equilibrium. Since in metal-poor
giants the \ion{Fe}{i} lines can deviate significantly from LTE
\citep[see e.g.][and references therein]{Mashonkina2011}, we decided
that the  isochrones and the colour-magnitude diagram  provide a more robust estimate
of the effective temperature.  
It
is not possible to measure lines on the linear part of the
curve-of-growth at the resolution of X-Shooter, thus one has no information 
on the microturbulent
velocity.  This parameter, which has to be introduced when using
one-dimensional static model atmospheres as done here, depends on the
atmospheric parameters.  We decided to adopt 2.0 \kms for the
microturbulent velocity because this was measured by
\citet{Cayrel} for HD\, 122563. This star has a \teff\ and \glog\
 similar to
our stars, but a lower metallicity. \citet{Monaco} derived a
calibration of the microturbulent velocity as a function of \teff\ and
log g  for metallicities in the range $-0.5$ to $-1.0$. Their calibration
implies microturbulent velocities $\approx 1.8$ \kms.

For Na and Ba, which are known to be strongly affected by deviations
from LTE, we derived the abundances from fitting line profiles with
theoretical profiles computed in NLTE with the code MULTI
\citep{Carlsson} as modified by \citet{Korotina} and \citet{Korotinb}.  For Na we
relied on the \ion{Na}{i} D doublet, the model atom used is the same
as in \citet{Andrievsky07}.  To derive the Ba abundances, we  used the
the 614.15\,nm \ion{Ba}{ii} line for both stars, and
for star Crater\,J113615-105244 we also used the 445.4\,nm line.
We used the model atom described in
\citet{Andrievsky09}.

The derived abundances for the two stars are given in Table\,\ref{abbo}.  
The effects of systematic errors due to different
assumptions on \teff, \glog,\ and microturbulent velocities for star
Crater\,J113613-105227 are clearly visible \relax in
Table\,\ref{error}.

As a check of our method, we also analysed\ the X-Shooter
spectrum of HD\,165195 with MyGIsFOS.  We assumed \teff=4500 K, \glog = 1.1, and
a microturbulence of 1.6\kms, as found by \citet{bonifacio99}.  The
derived [Fe/H] is $-2.08$, very close to what was found by
\citet[][--1.92]{bonifacio99}; this confirms the soundness of our method.

The derived abundances are consistent with the metallicity 
derived from the isochrone fits. 
The  PARSEC ischrones assume solar-scaled abundances.
Our detailed analysis reveals that the stars
are instead  $\alpha-$enhanced.
\citet{Salaris} have shown that for metal-poor stars an
$\alpha-$enhanced isochrone can be mimicked by a solar-scaled
one of higher metallicity. Using their prescription and
the mean of our measured [Fe/H] and [$\alpha$/Fe],
we find that our stars probably are compatible
with an isochrone that is only about 0.1\,dex more metal-rich
than the one shown in Fig. \ref{cmd}.
This is well within the errors of our abundance measurements
and the isochrone-fitting procedure.

\section{Discussion}
  
Our spectroscopic results confirm the metallicity estimated from the
photometry by \citet{Belokurov2014} and \citet{Laevens2014}. The
abundance pattern of the two stars does not show any signature that
distinguishes it from halo stars. The $\alpha$ elements are enhanced over
iron, and the iron peak elements follow iron.  Na is underabundant by 
roughly a factor of two with respect to iron in both stars,
as found in Galactic halo stars at this metallicity.

\begin{figure}
\begin{center}
\resizebox{\hsize}{!}{\includegraphics[clip=true]{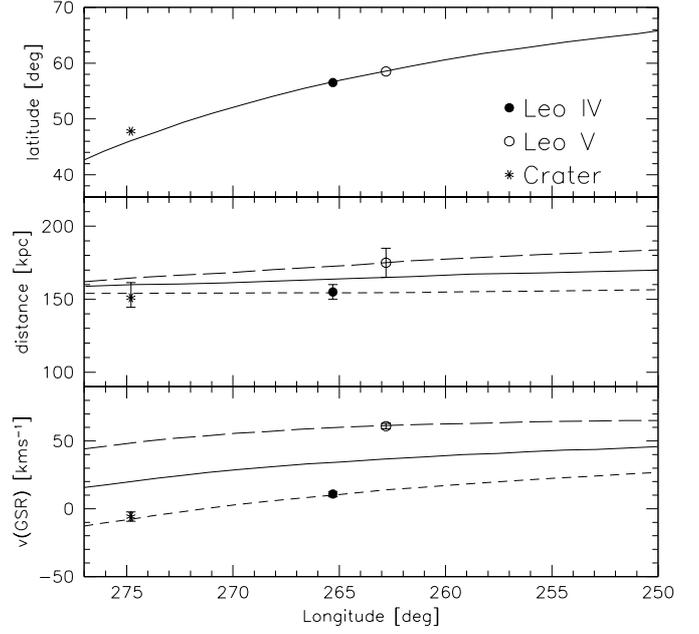}}
\end{center}
\caption[]{  Updated version of Fig. 7 of \citet{dejong}. 
The radial velocity (lower panel), distance (middle panel),
  and sky position (upper panel) of Crater (shown as a star) compared
  to those of Leo~IV (empty dot) and Leo~V (filled dot).  The lines are the orbits for Leo~IV
  and Leo~V, computed by \citet{dejong}, assuming common energy and
  angular momentum for the two galaxies, using Leo~IV and Leo~V (dotted
  line) as initial conditions, or their mean properties (dashed line),
  respectively.
\label{orbit}}
\end{figure}

The radial velocity of the system is very interesting (see Sect. \ref{radvel}).  
\citet{Belokurov2014} have already
pointed out that Crater, Leo~IV, and Leo~V all lie on a great circle
with a pole at ($\alpha,\delta$) = ($83^\circ.1,-5^\circ.3$).
\citet{dejong} suggested that Leo~IV and Leo~V are connected by a
``bridge of stars'' and could have been accreted by the Galaxy as a
pair.  \citet{Belokurov2014} suggested that Crater may in fact be a
member of the same group of galaxies.  Our measured radial velocity
and confirmed distance from the DES photometry strongly supports this
hypothesis. This is illustrated in Fig.\ref{orbit}, where we show the
positional and kinematical properties of the three galaxies, together
with the orbits computed by \citet{dejong} under the assumption that
Leo~IV and Leo~V have a common energy and angular momentum.

The association of Crater with the other two dwarf galaxies is evidence
that favours the interpretation of \citet{Belokurov2014}. It
might be argued that Crater is a GC associated with either
Leo~IV or Leo~V. This would make this whole system very exceptional
because with the exception of the Magellanic Clouds, the only Milky Way satellites
that harbour GCs are the two most massive dwarf
spheroidals: Sagittarius and Fornax.  The presence of a GC
system seems to be associated with a high mass; it would be
very unusual if Leo~IV or Leo~V, which have masses a factor of one thousand
lower than Sagittarius or Fornax \citep{McConnachie}, were indeed
possessed of such satellites.

The fact that the Galactocentric radial velocity of Crater is so low
implies that it is either close to peri-Galacton, or apo-Galacticon,
or on an orbit that is nearly circular. If the latter is true, or if the
star is close to peri-Galacticon, then Crater is spending most  of its time
at large distances ($> 150 $ kpc) from the Galactic centre, thus minimising
the stripping effect of the Galactic tidal field on the gas. 
This could explain why this dwarf galaxy has been able to retain gas
up to at least 2.2 Gyr ago and form a second generation of stars.

The radial velocities of the two stars observed by us differ by
10.2~\kms.   In Appendix \ref{gaus}  
we show that, ignoring errors on the radial velocity, 
this implies a lower limit on the dispersion
$\sigma_v > 3.7$ \kms with a 95\% confidence.

Globular clusters are known to have several tight relations that
connect their main
structural parameters. One of them is the fundamental plane, which
involves the central velocity dispersion, central density, and core
radius \citep{dj95}. Another is a relation that involves the total
luminosity and the velocity dispersion \citep{djm94}.  We can
therefore 
compare the velocity dispersion of Crater
with those of GCs. For a total luminosity of --5.5 \citep{Belokurov2014} or
--4.3 \citep{Laevens2014} for Crater, we expect a dispersion of
$2.5\div4.0$~\kms. For a proper comparison,   the ageing of
the whole stellar population of Crater needs to be taken into account. 
Crater is about  $5\div6$~Gyr younger
than the average population of GCs, which is of the order of
$12\div13$~Gyr.  In the comparison, Crater should migrate at fainter
magnitudes while not changing the velocity dispersion, which is barely
affected. 
Although the
comparison does not allow a strong
disctinction between Crater being a GC or a dwarf galaxy, 
it certainly does not favour it being a  GC.

Dwarf
galaxies have dynamical masses higher than the masses estimated
from their luminosities ($M_{dyn} >> M_*$), while for GCs the two
masses are of the same order of magnitude.  Our proposed lower limit
on the velocity dispersion can be combined with the radius and
distance of Crater to show that the dynamical mass of Crater is
$M_{Crater} > 1.5 M_{47Tuc}$ (if we assume the radius measured by
\citealt{Laevens2014}, $> 1.8M_{47Tuc}$ if we use the larger radius
measured by \citealt{Belokurov2014}, see Appendix for details).  When
coupled with the luminosities of Crater ($M_V = -4.3$
\citealt{Laevens2014}) and 47\,Tuc (--9.42 \citealt{Harris10}), this
immediately tells us that Crater must contain a significant fraction
of dark matter.  This consideration could, by itself, answer the
question and classify Crater as a dwarf galaxy.  Only two
stars have been measured, however, which weakens this argument
very much (e.g. one of
the two stars could be a binary or a radial velocity variable), and
more measurements are certainly required to draw a firm conclusion.

Stars bluer than the turn-off may be explained as a
younger population, which is another observation favouring the interpretation
of Crater as a dwarf galaxy.  Although it might be argued that these
stars could be blue stragglers, they would appear to be too numerous
for the cluster's luminosity.  \citet{Momany} showed that the logarithm of the ratio of the number of blue stragglers to
horizontal branch stars in GCs has a clear anti-correlation with cluster
luminosity. This anti-correlation is less tight for open clusters and
dwarf galaxies that occupy different regions in this plane. If we
count all stars to the blue of the turn-off in Crater as blue
stragglers and compare to the number of HB stars (see the two boxes
in Fig.~\ref{cmd}), this would place it in a region of Fig.~6.4
of \citet{Momany} that is mainly populated by dwarf galaxies and open
clusters, but not by GCs. Therefore even interpreting
the blue stars as blue stragglers would not support the notion that
Crater is a typical GC.

The photometry suggests that the metallicity of this
younger population has the same metallicity of the main population;
this is
expected.  Such low-mass galaxies (\citealt{Laevens2014}
estimated $6.8\times 10^3$M\sun) are expected to show a 
low metallicity dispersion. The reason is that they cannot effectively retain
supernova ejecta.

We summarise the four observations that all favour the interpretation
of Crater as a dwarf galaxy.
\begin{enumerate}
\item The likely association of Crater with Leo IV and Leo V,
on the basis of its radial velocity.
\item The population blueward of the turn-off in the
colour-magnitude diagram, which can be interpreted as a population
that is 4.5\,Gyr younger than the dominant population.
\item The lower limit to the velocity dispersion in the system
implies that Crater is more massive than 47 Tuc. Since it is also less luminous
than 47 Tuc, it must contain a significant fraction of dark matter.
\item Crater does not lie on the fundamental plane of GCs. 
\end{enumerate}

These observations are circumstantial, however,  and cannot rule
out the possibility that the system is a GC, if an
atypical one.
Nevertheless, from the information currently available, the dwarf
galaxy hypothesis seems more plausible.

\begin{acknowledgements}
We are grateful to the anonymous referee, who helped us to improve
the paper and pointed out to us that the almost horizontal track 
of the young population SGB does not follow the expected blue-straggler locus and that Crater must be near peri- or apo-galacticon
or on a nearly circular orbit, providing an argument in favour
of its ability to retain gas.
The project was funded by FONDATION MERAC.  PB, EC, PF, MS, FS, and RC
acknowledge support from the Programme National de Cosmologie et
Galaxies (PNCG) and Programme National de Physique Stellaire (PNPS) of
the Institut National de Sciences de l'Univers of CNRS.  Support for
L. S. was provided by Chile's Ministry of Economy, Development, and
Tourism's Millennium Science Initiative through grant IC120009,
awarded to The Millennium Institute of Astrophysics, MAS.
S.Z. acknowledges that this research was supported in part by the
National Science Foundation under Grant No. NSF PHY11-25915 and by
PRIN INAF 2014 - CRA 1.05.01.94.05 ``Star won't tell their ages to
GAIA''.  S.Z. warmly thank P. Ochner for useful observations
at the Asiago telescopes.
\end{acknowledgements}


\bibliographystyle{aa}

\Online

\begin{table*}
\caption{Detailed abundances.  For elements for which several lines
are measured, the quoted  error is the line-to-line scatter, except for \ion{Na}{i}. 
For Crater\,J113613-105227 the
two lines of the doublet where fitted together,  and
for  Crater\,J113615-105244 only the D2 line was fitted.
The quoted errors are estimated from the
goodness-of-fit. For the elements for which only one line was measured, we provide
no estimate of the error, but recommend to take the error estimate provided for
\ion{Fe}{i}, since for both stars this is the element with the largest number of measured
lines, and the line-to-line scatter can be interpreted as mainly due to the noise
in the spectrum.}
\label{abbo}
\begin{center}
\begin{tabular}{lrcrrrrrrrrrrr}
\noalign{\smallskip}\hline\noalign{\smallskip}
 Ion  & A(X)\sun  & N &  A(X)    &   [X/H] & $\sigma$  &  [X/Fe] & $\sigma$  & N &  A(X)  &  [X/H]  & $\sigma$  &   [X/Fe] & $\sigma$  \\
     &       &   \multicolumn{6}{c}{J113613-105227}                 & \multicolumn{6}{c}{J113615-105244}                  \\
\noalign{\smallskip}\hline
\noalign{\smallskip}
\ion{Na}{i}  &   6.30 &  2 & 4.35    &$ -1.95$ & 0.20 &  $-0.22$    &0.36 & 2  &  4.35 & $-1.95$ & 0.20 & $-0.28$  & 0.36 \\
\ion{Mg}{i}  &   7.54 &  2 & $ 6.12$ & $-1.42$ & 0.03 &  $ 0.31$ & 0.26 &   2  &  6.22 & $-1.32$ & 0.42 & $ 0.35$  & 0.51 \\
\ion{Ca}{i}  &   6.33 &  4 & $ 4.96$ & $-1.37$ & 0.23 &  $ 0.36$ & 0.35 &   4  &  5.10 & $-1.23$ & 0.11 & $ 0.44$  & 0.30 \\
\ion{Ti}{ii} &   4.90 &  3 & $ 3.36$ & $-1.54$ & 0.19 &  $ 0.19$ & 0.20 &   2  &  3.21 & $-1.69$ & 0.20 & $-0.02$  & 0.34   \\
\ion{V}{i}   &   4.00 &  1 & $ 2.11$ & $-1.89$ &      &  $-0.16$ &      &   1  &  2.23 & $-1.77$ &      & $-0.10$  &        \\
\ion{Cr}{ii} &   5.64 &  2 & $ 3.82$ & $-1.82$ & 0.03 &  $-0.09$ & 0.26 &   1  &  4.40 & $-1.24$ &      & $ 0.43$  &        \\
\ion{Mn}{i}  &   5.37 &  2 & $ 3.74$ & $-1.63$ & 0.09 &  $ 0.10$ & 0.28 &      &       & $     $ &      & $     $  &      \\
\ion{Fe}{i}  &   7.52 & 58 & $ 5.79$ & $-1.73$ & 0.26 &  $ 0.00$ &      &  32  &  5.85 & $-1.67$ & 0.28 & $ 0.00$  &      \\
\ion{Fe}{ii} &   7.52 &  2 & $ 5.67$ & $-1.85$ & 0.05 &  $ 0.00$ &      &   1  &  5.97 & $-1.55$ &      & $ 0.00$  &        \\
\ion{Ni}{i}  &   6.23 &  2 & $ 4.55$ & $-1.68$ & 0.18 &  $ 0.05$ & 0.31 &   1  &  4.62 & $-1.61$ &      & $ 0.05$          \\
\ion{Ba}{ii} &   2.17 &  1 & $ 0.49 $ & $-1.68$ &      &   $-0.01$&      &  2  & $ 0.43$ & $-1.74$ & 0.20    & $-0.07$  0.34    \\
\noalign{\smallskip}\hline\noalign{\smallskip}
\end{tabular}
\end{center}
\end{table*}


\begin{table}
\caption{Variations in stellar parameters for different assumptions on \teff, \glog, and microturbulent velocity.}
\label{error}
\begin{center}
\begin{tabular}{rrrrr}
\hline
\noalign{\smallskip}
 T${\rm eff}$ & log\,g & $\xi$ &[Fe/H] & [$\alpha$/Fe] \\
     K        & [cgs]  & \kms  & dex   &  dex \\
\noalign{\smallskip}
\hline
\noalign{\smallskip}
\noalign{\smallskip}
 \multicolumn{5}{c}{J113613-105227} \\
\noalign{\smallskip}
\hline
 \multicolumn{5}{c}{Different assumed \teff} \\
\hline
\noalign{\smallskip}
 4475& 1.24 & 2.0 & $-1.89$ &  0.34 \\
 4575 & 1.24 & 2.0 & $-1.73$ &  0.34 \\
 4675 & 1.24 & 2.0 & $-1.60$ &  0.36 \\
\noalign{\smallskip}             
\hline
 \multicolumn{5}{c}{Different assumed \glog}\\
\hline                           
\noalign{\smallskip}             
 4575 & 1.04 & 2.0 & $-1.73$ &  0.39 \\
 4575 & 1.24 & 2.0 & $-1.73$ &  0.34 \\
 4575 & 1.44 & 2.0 & $-1.74$ &  0.30 \\
\noalign{\smallskip}             
\hline
 \multicolumn{5}{c}{Different assumed $\xi$}\\
\hline                           
\noalign{\smallskip}             
 4575 & 1.24 & 1.6 & $-1.54$ &  0.23 \\
 4575 & 1.24 & 1.8 & $-1.63$ &  0.28 \\
 4575 & 1.24 & 2.0 & $-1.73$ &  0.31 \\
 4575 & 1.24 & 2.2 & $-1.81$ &  0.37 \\
 4575 & 1.24 & 2.4 & $-1.90$ &  0.42 \\
\noalign{\smallskip}             
\hline
 \multicolumn{5}{c}{\glog\ from Fe ionisation equilibrium}\\
\hline                           
\noalign{\smallskip}             
 4575 & 1.62 & 2.0 & $-1.75$ &  0.26 \\
\noalign{\smallskip}
\hline
\end{tabular}
\end{center}
\end{table}

\appendix
\section{lower limit to the radial velocity dispersion\label{gaus}}
The difference of two normal random variables with means $\mu_1,\mu_2)$,
and standard deviations $\sigma_1,\sigma_2$
 follows a
normal distribution, with $\sigma = \sqrt{(\sigma_1^2 + \sigma_2^2)}$
and mean $\mu_1-\mu_2$.  
Assuming that the velocity
distribution is normal and ignoring the measurement errors, 
the velocity difference between two stars is expected to be normally distributed
with a mean of 0 \kms and  a standard deviation
of $\sqrt{2}\sigma_v$, where $\sigma_v$ is the standard deviation
of the original distribution. 
Using the table on page 253 of \citet{Bevington}, which provides the integral
of a Gaussian distribution, we derive that the integral between
$-1.96 \le z \le +1.96$ is 0.95, where $z = \frac{|x-\mu|}{\sigma}$
and $\mu$ is the mean. In our case, x=10.2 \kms, $\mu =$ 0 \kms
and $\sigma = \sqrt{2}\sigma_v$; this implies that $z=1.96$ corresponds $\sigma_v=3.7$\kms. 
In other words,  there is a 95\% probability that the parent velocity distribution
has a $\sigma> 3.7$~\kms.  
If we take into account the errors, no conclusion can be reached
because the measured 
radial velocities of the two stars are also consistent with a zero radial velocity
dispersion.

\section{lower limit to the dynamical mass of Crater}

To derive a lower limit, we assumed that Crater and the globular cluster
47\,Tuc are in dynamical equilibrium and that the virial theorem applies.
For this system, the mass, $M, $ can be expressed in terms of its velocity
dispersion, $\sigma$, and exponential radius, $r_h$ \citep[see e.g.][]{Spitzer87} :
\begin{equation}
M ={{\sigma^2 r_h}\over{0.4 G}}
,\end{equation}

where $G$ is the gravitational constant.
We thus may write
\begin{equation}
{M_{Crater}\over M_{47TUC}} = {\sigma_{Crater}^2\over\sigma_{47Tuc}^2}{r_{hCrater}\over r_{h47Tuc}}
.\end{equation}

Because we only have a lower limit for $\sigma_{Crater}$, this becomes

\begin{equation}
{M_{Crater}\over M_{47TUC}} > {\sigma_{Crater}^2\over\sigma_{47Tuc}^2}{r_{hCrater}\over r_{h47Tuc}}
.\end{equation}

We assumed for Crater $\sigma_{Crater}> 3.7 $\kms, $r_{hCrater} = 0.47'$ \citep{Laevens2014}
or $r_{hCrater} = 0.6'$ \citep{Belokurov2014} and a distance of 145 kpc \citep{Laevens2014}.
For 47\,Tuc we assumed $\sigma_{47Tuc} = 7$ \kms\ a median value from \citet{Kucinskas},
$r_{h47Tuc} = 174 ''$ \citep{Trager93}, and a distance of 4.5 kpc \citep{Harris10}.

With these values, we obtain ${M_{Crater}\over M_{47TUC}} > 1.5$ for the radius measured by
\citet{Laevens2014} and ${M_{Crater}\over M_{47TUC}} \ga 1.9$ for the radius measured
by \citet{Belokurov2014}. Either way, it is clear that Crater is more massive (dynamically)
than 47 Tuc. Even disregarding the velocity dispersions, it might have been suspected that
this is the case because the radius of Crater is larger than that of 47 Tuc.
Of course, if Crater is not in dynamical equlibrium, this reasoning does not hold.

\end{document}